\documentclass[12pt]{article}
\usepackage{amssymb,amsmath,epsfig}
\allowdisplaybreaks

\begin{document}

\title{\bf Pulsation of black holes}

\author{\small Changjun Gao,\thanks{gaocj@bao.ac.cn}
\small ~~Youjun Lu\thanks{luyj@bao.ac.cn}\\
\small\it Key Laboratory for Computational
\small\it Astrophysics, \small\it \\ \small\it National Astronomical
\small\it Observatories, \small\it \\ \small\it Chinese Academy of Sciences, Beijing,
 \small\it 100012, China\\
 \small\it School of Astronomy and Space Sciences,  \small\it \\ \small\it University of Chinese Academy of Sciences,  \small\it\\ \small\it No. 19A, Yuquan Road, Beijing 100049, China\\
 \small You-Gen Shen\thanks{ygshen@center.shao.ac.cn}\\
\small\it Shanghai Astronomical Observatory,\\ \small\it Chinese
\small\it Academy of Sciences, Shanghai 200030, China\\
\small Valerio Faraoni\thanks{vfaraoni@ubishops.ca}\\
 \small\it Department of Physics and Astronomy, Bishop's
University,
\\ \small\it 2600 College Street, Sherbrooke, Qu\'ebec, Canada J1M~1Z7}

\date{}
\maketitle

\begin{abstract}

The Hawking-Penrose singularity theorem states that a
singularity forms inside a black hole in general
relativity. To remove this singularity one must resort to a
more fundamental theory. Using  a corrected dynamical
equation arising in
loop quantum cosmology and braneworld models, we study the
gravitational collapse of a perfect fluid sphere with
a rather general equation of state. In the frame of an
observer
comoving with this fluid, the sphere pulsates between a
maximum and a minimum size, avoiding the singularity. The
exterior geometry is also constructed. There are usually
an outer and an inner apparent horizon,
resembling
the Reissner-Nordstr\"om situation. For a
distant observer the {horizon} crossing occurs in an
infinite time and the pulsations of the black hole quantum
``beating heart'' are completely unobservable. However, it may be observable
if the black hole is not spherical symmetric and radiates gravitational
wave due to the quadrupole moment, if any.

\end{abstract}

\clearpage

\section{Introduction}
\label{Sec:1}

General relativity predicts singularities at
the center of black holes. This prediction is
reinforced by the Hawking-Penrose singularity theorem
\cite{haw:1970}, which applies if the following four
physical assumptions are made: (i)~the Einstein equations
hold, (ii)~the strong energy condition holds, (iii)~there
are no closed timelike curves and, (iv)~every timelike or
null geodesic enters a region where the curvature is not
specially aligned with the geodesics. However,
singularities signal the breakdown of general relativity
and it is generally believed that they will be
removed in a more fundamental
theory of quantum gravity. Hence, while general
relativity provides a highly successful description of
gravity from  sub-millimeter to cosmic scales, it is
expected to break down at scales around the Planck
length thus constituting, at best, an
approximation to a truly fundamental theory of gravity.

Currently, there is no universally accepted theory of quantum gravity.
Two of the main competitors to this role are string theory
\cite{sch:2003} and loop quantum gravity (LQG) \cite{lqg:2004}. Hence,
it is important to explore the gravitational collapse problem in the
framework of these theories. Here we revisit the pioneering work on
classical gravitational collapse by Oppenheimer and Snyder
\cite{open:1939}, but we correct it according to the tenets of loop
quantum cosmology (LQC) and braneworld models. Many interesting studies
on the resolution of black hole singularities have been carried out
directly in the framework of LQG or in polymer quantization (which is an
offshoot of LQG \cite{AFW03}) \cite{Boe:2007, mod:2006,ale:2015,ol:2017,
cam:2007, boj:2005, mod:2008, chiou:2008, boj:2008,
Rod:2008,TippettHusain11, ModestoPRD04, ZiprickKunstatter,
PeltolaKunstatter,bambi,liu,mala}, to which the present paper intends to
contribute. Specifically, by using the corrected dynamical equation of
LQC or braneworld models, we study the gravitational collapse of a
perfect fluid sphere with a rather general equation of state. {\em In
the context of LQC, our study amounts to an approximate analysis of
effective dynamics, which is reflected in approximating Eq.~(\ref{modifiedFriedmann1}) below
with Eq.~(\ref{modifiedFriedmann}).  }

It is found that, in the comoving frame, the sphere does
not collapse to a
singularity but instead pulsates between a maximum and
a minimum size, avoiding the singularity.  We propose a
method to construct the exterior spacetime.
There are usually two {apparent} horizons in this
exterior geometry,
resembling the situation for the Reissner-Nordstr\"om black
hole. In the
frame of an observer at spatial infinity, the collapsing
fluid crosses the {outer} horizon in an infinite
coordinate
time and the pulsations are completely unobservable.
Borrowing terminology now recurrent in the literature on
astrophysical black holes and their X-ray periodicities, the
pulsating core inside the horizon can be described as the
``beating heart'' of the quantum-corrected black hole.

This paper is organized as follows. In Sec.~\ref{Sec:2} we
consider the gravitational collapse of a perfect fluid
sphere with the help of {\bf a} dynamical equation stemming
from LQC and braneworld models. In Sec.~\ref{Sec:3} we
perform a coordinate transformation to Schwarzschild-like
coordinates in order to extend the solution outside the
sphere and to spatial infinity. In Sec.~\ref{Sec:4} we
verify that the interior geometry matches the exterior one
{continuously} according to the minimal matching
conditions of
general
relativity.  Sec.~\ref{Sec:5} contains a discussion and the
conclusion. We adopt units in which $G=c=\hbar=1$, the
metric signature $-,\ +,\ +,\ +$, and we follow the
notation of Ref.~\cite{Waldbook}.

\section{Pulsating fluid sphere in the comoving frame}
\label{Sec:2}

In principle, a variety of different interiors can be
conceived for a perfect fluid sphere with surface $\Sigma$.
Let us  focus on
the simplest one, {\em i.e.}, an interior which
is spatially homogeneous and isotropic everywhere except
at its surface $\Sigma$, that is, an interior geometry
which is locally a portion of a closed
Friedmann-Lema\^itre-Robertson-Walker (FLRW) universe
\cite{haw:1973} with line element
\begin{equation}
ds^2=-dt^2+a^2\left(\frac{dr^2}{1-kr^2}
+r^2 d\Omega_{(2)}^2  \right)\,.\label{1}
\end{equation}
$a(t)$ is the scale factor associated with the
(time-dependent) proper
radius $a(t)r$ of the fluid sphere, $k$ is a positive
constant, and $d\Omega_{(2)}^2=d\theta^2 +\sin^2 \theta \,
d\varphi^2$ is the line element on the unit 2-sphere.
The Friedmann equation is
\begin{equation}
{\dot{a}^2}=\frac{8\pi}{3}\rho a^2-k\,,\label{Friedmann}
\end{equation}
where the overdot denotes differentiation with respect to
the comoving time $t$.
The energy densities of  pressureless matter (dust),
radiation, and stiff matter are
\begin{eqnarray}
&&\rho=\frac{\rho_p}{a^3}\,, \ \
\rho=\frac{\rho_r}{a^4}\,,\  \  \ \textrm{and} \ \ \
\rho=\frac{\rho_s}{a^6}\,,
\end{eqnarray}
respectively. One can generically model these fluids
as
\begin{eqnarray}
\rho=\frac{\rho_f}{a^{f+2}}\,,\ \ \ \textrm{ with}\ \ \
f=1, 2,  4 \,,
\end{eqnarray}
where $\rho_f$ is a constant. Assuming that the fluid
sphere is initially at rest at $t=0$ with initial
radius $a_{0}$, we have
\begin{equation}
\dot{a} (0)=0 \,, \label{2}
\end{equation}
and
\begin{equation}
k=\frac{8\pi}{3}\cdot\frac{\rho_f}{a_0^f}\,.
\end{equation}
Then Eq.~(\ref{Friedmann}) can be written as
\begin{equation}
{\dot{a}}=-\sqrt{k\left[\left(\frac{a_0}{a}\right)^{f}
-1\right]}\,,
\end{equation}
and its  solution can be expressed in
terms of
the hypergeometric function as
\begin{equation}
t=-\frac{i a}{\sqrt{k}}
\cdot\textrm{hypergeom} \left(\left[\frac{1}{2},
-\frac{1}{f}\right],\left[
1-\frac{1}{f}\right],\frac{a_0^f}{a^f}\right)+\mbox{const.}\,,
\end{equation}
where $i$ is the imaginary unit. This
solution reveals that, classically, the fluid sphere
collapses from finite initial radius $a_0$ to a state of
vanishing radius $a=0$ in a finite proper time of the order
\begin{equation} t\sim\frac{a_0}{\sqrt{k}}\,.
\end{equation} The final classical state of the fluid
sphere has infinite density and curvature and a singularity
is formed: if $f=1$ this is the well known result of
Oppenheimer and
Snyder for dust collapse \cite{open:1939}. The singularity
is shown to occur
also in more general situations by the singularity theorem
of Hawking and Penrose \cite{haw:1970} and its
generalizations \cite{Waldbook}. Singularities
signal the breakdown of Einstein's theory of gravity and
should be removed in a more fundamental theory of gravity,
such as LQG or string theory. In the following we discuss a
different picture of this gravitational collapse situation
obtained by introducing quantum gravity corrections.

For the evolution of the collapsing fluid sphere,  both LQC
\cite{lqc:001,lqc:002,lqc:003,lqc:0031,lqc:0032,lqc:004,para:14,lqc:005,lqc:006,
Bojowaldbook} and braneworld models \cite{bw:001,bw:002,bw:003} modify
the Friedmann equation to
\begin{equation}
{\dot{a}^2}=\frac{8\pi}{3}\rho\left(1-\frac{\rho}{
\rho_{cr}}\right)a^2\;, \label{nok}
\end{equation}
with vanishing of $k$.  $\rho_{cr}$ is a critical density which is generally
assumed to be of the order of the Planck energy density
$\rho_{pl}$, for example,
$\rho_{cr}=\sqrt{3}/32\pi^2\gamma^3\rho_{pl}\simeq 0.41\rho_{pl}$
\cite{lqc:007}, where $\gamma=0.2375$
is the dimensionless Barbero-Immirzi parameter. {However,
{ if $k$ is non-zero}, braneworld models
\cite{bw:001,bw:002,bw:003} and LQC \cite{lqc:003,lqc:0031,lqc:0032} give
 \begin{equation}
{\dot{a}^2}=\frac{8\pi}{3}\rho\left(1-\frac{\rho}{
\rho_{cr}}\right)a^2-k\;,  \label{modifiedFriedmann}
\end{equation}
and
\begin{equation}
{\dot{a}^2}=\frac{8\pi}{3}\left(\rho-\rho_1\right)\left(\frac{\rho_2}{\rho_{cr}}-\frac{\rho}{
\rho_{cr}}\right)a^2\;,  \label{modifiedFriedmann1}
\end{equation}
respectively. Here $\rho_1$ and $\rho_2$ are related to $k$ as follows
\begin{equation}
\rho_1=-k\chi\rho_{cr}\;,\ \ \ \ \rho_2=\rho_{cr}\left(1-k\chi\right)\;.
\end{equation}
It is apparent { that}
Eq.~(\ref{modifiedFriedmann}) and
Eq.~(\ref{modifiedFriedmann1}) are different in the
presence of { a non-zero} $k$.
Eq.~(\ref{modifiedFriedmann1})
is a fully non-perturbative modification coming from the
nonlocal nature of the field strength in the quantum
Hamiltonian constraint. In fact, Ref.~\cite{lqc:0071}
clearly shows that, in order to obtain this equation from
an action framework, infinite number of terms in curvature
invariants are required. On the other hand,
Eq.~(\ref{modifiedFriedmann}) is derived, in braneworld
models, {as} a tree level effect and with the
assumption of { an} extra time dimension
\cite{bw:001,bw:002}. Numerical evidence
shows that the effective equation Eq.~(\ref{nok}) provides
an excellent approximation to the full dynamics of sharply
peaked states, including at the bounce point where quantum
gravity effects are strongest
\cite{lqc:004,rev:2,piter:14,rev:3}. \emph{We note that in the limit, $\chi\rightarrow 0$, we obtain
\begin{equation}
\frac{\rho_2}{\rho_{cr}}\rightarrow 1\;, \ \ \ \ \frac{\rho_1}{\rho_{cr}}\rightarrow 0\;,\ \ \ \ \rho_1\rightarrow\frac{k}{a^2}\;.
\end{equation}
Thus we go back to Eq.~(11) from Eq.~(12). (For the last equation, see Eq.~(31) in Ref.~[28]).} In this paper, {
motivated by calculational simplicity}, we
shall use Eq.~(\ref{modifiedFriedmann}) to investigate the
gravitational collapse problem. }
\begin{figure}[h]
\begin{center}
\includegraphics[width=9cm]{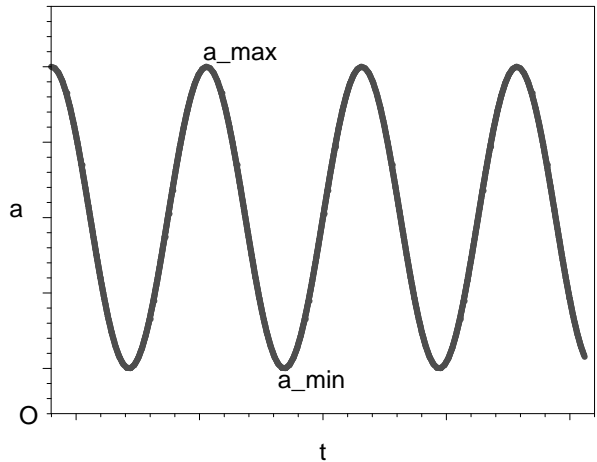}
\caption{The evolution of the radius of the
fluid sphere. This sphere oscillates between
a maximum  and a minimum radius and the
classical singularity
at $a=0$ is never reached.}\label{fig:1}
\end{center}
\end{figure}

In Fig.~\ref{fig:1} we plot the evolution of
the radius of the fluid sphere with
$\rho=\rho_f/ a^{f+2} $. The radius of
the fluid sphere oscillates between a  maximum
$a_\text{max}=a_0$ and a minimum $a_\text{min}$ and the
singularity does not form. The reason for this behavior
can be understood as follows.
According to the modified
equation~(\ref{modifiedFriedmann}), the
energy density continually increases as the
sphere contracts, for as long as the $k$-term in the right
hand
side of this equation can
be neglected. As the density approaches the critical
density the collapse stops and, thereafter, the sphere
begins to expand. As it reaches the maximum
radius, the expansion stops and the sphere begins
to collapse again.

The scenario described is the picture observed by a
comoving observer. Since the singularity is not formed, the
question arises of whether there is a black hole horizon or
not. To answer this question, it is necessary to consider
the frame of an observer located at spatial infinity.

\section{Observer at spatial infinity}
\label{Sec:3}

Here we extend the spacetime of Eq.~(\ref{1}) across the
surface $\Sigma$ of the sphere and to spatial
infinity. The angular coordinates $\theta$ and $\varphi$
are the same in both regions. Let us begin by rewriting
the line element in different coordinates. By
introducing the variable
\begin{equation}
r \equiv \frac{x}{a}\,, \label{x}
\end{equation}
the line element~(\ref{1}) becomes
\begin{eqnarray}
ds^2 & = & -\left(1-\frac{x^2H^2}{1-kr^2}\right)dt^2
-\frac{2xH}{1-kr^2} \, dt \, dx \nonumber\\
&&\nonumber\\
& \, &+\left(1-kr^2\right)^{-1}dx^2
+x^2d\Omega_{(2)}^2\,,\label{lineelementA}
\end{eqnarray}
where
\begin{eqnarray}
H\equiv\frac{\dot{a}}{a}
\end{eqnarray}
is the usual Hubble parameter. In order to eliminate the
cross-term  $dt \, dx$ in the line
element~(\ref{lineelementA}), we
replace the
comoving time $t$ with another time coordinate
$\bar{t}$. To determine $\bar{t}$ we employ the
``integrating factor method'' \cite{wein:1972}, that is, we
set
\begin{eqnarray}
d\bar{t}=\eta^{-1}\left[-\left(1-\frac{x^2H^2}{1-kr^2}
\right)dt-\frac{xH}{1-kr^2} \,
dx\right]\,.\label{integrating}
\end{eqnarray}
Here $\eta(t,\ x)$ is an integrating factor
making $d\bar{t}$ a perfect differential, which is achieved
if $\eta $ satisfies the equation
\begin{eqnarray}
\partial_{x}\left[\eta^{-1}\left(1-\frac{x^2H^2}{
1-kr^2}\right)\right]=\partial_t\left( \eta^{-1}
\frac{xH}{1-kr^2}\right) \,. \label{16}
\end{eqnarray}
The solution of this equation is
\begin{eqnarray}
&&\eta=a^2H {F}\left(u\right)\; \ \ \
u \equiv \mbox{e}^{-2k{W}}
\left(1-kr^2\right)\,\nonumber\\
&&\nonumber\\
&&{W} \equiv\int\frac{dt}{a^2H} \,
\end{eqnarray}
where ${F}(u)$ is an arbitrary function of $u$
(which makes it clear that the integrating factor is not
unique). The line element~(\ref{lineelementA}) then becomes \cite{wang:2013}
\begin{eqnarray}
ds^2&=&-\eta^2\left(1-\frac{x^2H^2}{1-kr^2}
\right)^{-1}d\bar{t}^2+x^2d\Omega_{(2)}^2\nonumber\\
&&\nonumber\\
&&+\left( 1-kr^2-x^2H^2\right)^{-1}dx^2\,.
\end{eqnarray}
By defining
\begin{eqnarray}
F^2\equiv \eta^2\frac{1-kr^2}{\left(
1-kr^2-x^2H^2\right)^{2}}\,, \label{18}
\end{eqnarray}
it is
\begin{eqnarray}
ds^2&=&-F^2\left[1-x^2\left(H^2+\frac{k}{a^2}
\right)\right]d\bar{t}^2+x^2d\Omega_{(2)}^2\nonumber\\
&&\nonumber\\
&\, &+\left[1-x^2\left(H^2+\frac{k}{a^2}\right)
\right]^{-1}dx^2\,.\label{interior}
\end{eqnarray}
We now want to use the $\left( \bar{t}, x, \theta, \varphi
\right)$ coordinate system to extend the geometry
determined by \emph{quantum gravity} outside the fluid
sphere. The resulting metric is interpreted as a \emph{quantum gravity}
geometry with a fluid also outside the
surface $\Sigma$ of this sphere. {If the matter density is much smaller than the Planck density, the quantum correction in Eq.~(\ref{modifiedFriedmann}) becomes negligible and the
geometry reduces to a fluid solution of the
Einstein equations.} Using the modified
Friedmann equation~(\ref{modifiedFriedmann}), the line
element assumes the form
\begin{eqnarray}
ds^2 &=& -F^2\left[1-\frac{8\pi}{3}\rho\left(
1-\frac{\rho}{\rho_{cr}}\right)x^2\right]
d\bar{t}^2+x^2d\Omega_{(2)}^2\nonumber\\
&&\nonumber\\
&&+\left[1-\frac{8\pi}{3}\rho\left(1-\frac{\rho}{
\rho_{cr}}\right)x^2\right]^{-1}dx^2\,.
\end{eqnarray}
Using now the scaling of the perfect fluid density
$\rho=\rho_f
/ a^{f+2}$, one obtains
\begin{eqnarray}
ds^2&=&-F^2\left[1-\frac{8\pi}{3}\frac{\rho_f
r^{f+2}}{x^f}\left(1-\frac{\rho_f r^{f+2}}{
\rho_{cr}x^{f+2}}\right)\right]d\bar{t}^2\nonumber\\
&&\nonumber\\
&\, &  +\left[1-\frac{8\pi}{3}\frac{\rho_f
r^{f+2}}{x^f}\left(
1-\frac{\rho_f
r^{f+2}}{\rho_{cr}x^{f+2}}\right)\right]^{-1}
dx^2\nonumber\\
&&\nonumber\\
&\, & +x^2d\Omega_{(2)}^2\,. \label{21}
\end{eqnarray}
Here the metric components must be understood as
functions of $\bar{t}$ and $x$ obtained by solving
Eqs.~(\ref{x}), (\ref{integrating}), (\ref{16}),
and~(\ref{18}). This metric continues the
interior (FLRW) geometry of the fluid sphere across
$\Sigma$ to the
exterior and to
spatial infinity, provided that suitable junction
conditions (discussed below) are satisfied on $\Sigma$. In
order to understand completely the
behavior of the quantum-corrected fluid sphere, one should
determine also the spacetime structure outside the sphere.
What is this exterior geometry? Is it the Schwarzschild
spacetime? The answer is no, for the following reason.
At the surface $\Sigma$ (a timelike three-dimensional
world tube which has the sphere as its intersection with
any constant time slice), the exterior and interior
geometries
must match {continuously} \cite{MTW:1973}. According
to the
Israel junction conditions, if the match cannot
be achieved, then there is a layer of material on the
surface $\Sigma$ of the sphere. The relation between the
jump of the
stress-energy tensor at the surface and its intrinsic
geometry is regulated by field equations (in general
relativity, the jump of the Einstein tensor is proportional
to the jump of the stress-energy tensor according to the
Einstein equations) \cite{Israel, Poisson}. In our case,
however, no layer of
material is present on $\Sigma$  and the
minimal junction conditions required in any theory of
gravity, that is {continuous}  matching of induced
3-metrics and extrinsic curvatures, are imposed.
We note that similar points are clarified in general terms
{in \cite{barcelo:2016}}.

In the following we obtain the exterior spacetime
using a ``surface trick'' (this method is described in
Appendix~\ref{AppendixA} for general static and spherically
symmetric spacetimes).  We verify that the exterior
geometry matches {continuously} the interior FLRW
geometry at the surface of the fluid sphere.

At the surface of the sphere $r=r_0$ (where, since $r$ is
a comoving coordinate, $r_0$ is a
constant \cite{MTW:1973}), Eq.~(\ref{21}) yields
\begin{eqnarray}
ds^2&=&-\left[1-\frac{2M_f}{x^{f}}\left(
1-\frac{l^{f+2}}{x^{f+2}}\right)\right]
d{\tilde{t}}^2+x^2d\Omega_{(2)}^2\nonumber\\
&&\nonumber\\
&&+\left[1-\frac{2M_f}{x^{f}}\left(
1-\frac{l^{f+2}}{x^{f+2}}\right)\right]^{-1}dx^2\,,
\label{22}
\end{eqnarray}
and
\begin{eqnarray}
M_f \equiv \frac{4\pi}{3} \, \rho_f \, r_0^{f+2} \,,\ \ \ \
\ \ \ \ \ \ \ l^{f+2} \equiv \frac{\rho_f \,
r_0^{f+2}}{\rho_{cr}} \,.
\label{23}
\end{eqnarray}
Here $\tilde{t}$ is defined as
\begin{equation}
\tilde{t} = \int F\left(\bar{t},\ r_0\right)d\bar{t}\,.
\end{equation}

When $f=1$ (describing a  pressureless dust), we recognize
${M}_1$ as the fluid energy enclosed by
the sphere. The
${l}$-terms in the line element are generated by  the
LQC/braneworld correction. Eq.~(\ref{22}) describes
a 2-parameter family of static and spherically symmetric
spacetimes spanned by the parameters $M_f$ and
$l$. When
$f=1$ and $l\ll x$, Eq.~(\ref{22}) reduces to the
Schwarzschild  solution of the vacuum Einstein equations.
At first glance, the
metric~(\ref{22}) looks like a special case of the Kiselev
metric \cite{kis:2003}, but differs from it in the sign and
the $x$-dependence of the $l$-term.

Does the exterior metric~(\ref{22}) satisfy the
Einstein equations outside the sphere? Clearly, it does
not satisfy the vacuum Einstein equations because the Birkhoff
theorem of general relativity guarantees that the
unique vacuum, spherical, and asymptotically flat solution
is the Schwarzschild one. Furthermore, the Schwarzschild
metric can not match {continuously} the interior
FLRW geometry for $f=2,4$, unless
there is
a layer of material on the surface of the sphere, which
introduces some arbitrariness of choice that is best
avoided. In order to match {continuously} exterior
and interior
geometry without matter layers on the surface $\Sigma$ of
the
sphere, the exterior solution must be given by
Eq.~(\ref{22}). The interior FLRW geometry is not derived
from
the Einstein equations but from the braneworld modified Friedmann
equation. Therefore, one
should in principle construct the exterior solution from
the equations of braneworld model. However, in general, one
expects that  $l/x\ll 1$ in the exterior and that the quantum
correction ${(l/x)}^{f+2}$ can be safely neglected outside
the sphere.

{We note that, actually, considering the trace anomaly of quantum fluctuations,
Abedi and Arfaei \cite{abe:2016} have obtained the quantum corrected metric Eq.~[\ref{22}] for $f=1$}.

The equation locating the {apparent} horizons (when
they exist) is
\begin{eqnarray}
1-\frac{2M_f}{x^{f}}\left(1-\frac{l^{f+2}}{x^{f+2}}\right)=0\,,
\end{eqnarray}
or
\begin{eqnarray}
1-\frac{8\pi}{3}\frac{\rho_f \, r_0^{f+2}}{
x^f}\left(1-\frac{\rho_f \, r_0^{f+2}}{\rho_{cr}
\, x^{f+2}}\right)=0\,.
\end{eqnarray}
In general, this equation has two positive solutions,
which means that there exist two {apparent} horizons
with radii
 $x_+$ and $x_{-}$. $x_{+}$ denotes the outer
{apparent}
horizon, a surface of infinite redshift, while
$x_{-}$ denotes an inner {apparent} horizon
\cite{chen:2015}. In particular,
when $f=2$, then $x_{+}$ and $x_{-}$ can be expressed as
\begin{eqnarray}
x_{+} &=& \left( \frac{2M_2}{3} \right)^{1/2}
\Bigg( 2\cos\mathfrak{s}+1 \Bigg)^{ 1/2}\,,\\
&& \nonumber\\
x_{-}&=&\left(\frac{2M_2}{3} \right)^{1/2}
\left[2\cos\left(\mathfrak{s}+\frac{4}{3}\pi\right)
+1\right]^{1/2}\,,
\end{eqnarray}
where
\begin{eqnarray}
\mathfrak{s}\equiv \frac{1}{3}\arccos\left({1-\frac{27
l^4}{8M_2^2}}\right)\,.
\end{eqnarray}
When
\begin{eqnarray}
1-\frac{27 l^4}{16M_2^2}=0\;,
\end{eqnarray}
the two {apparent} horizons coincide. When $l^4\ll
M_2^2$, then
$x_+=\sqrt{2M_2}$ gives the {outer} horizon and
$x_{-}=0$ gives the singularity. The causal structure of
the exterior spacetime is similar to that of the
Reissner-Nordstr\"om solution of the Einstein equations:
the outer horizon is an {apparent}  horizon, the
inner {apparent} horizon {resembles}
a Cauchy horizon \cite{chen:2015}, and
the singularity is timelike. Therefore, the solution can
also be used to describe a wormhole or
Einstein-Rosen bridge connecting two asymptotically flat
spacetimes.

Now a question arises: in the frame of an observer at
spatial  infinity, will the fluid sphere oscillate?
Is the surface of this sphere inside or outside the
{apparent} horizons? To answer these questions, we now turn to
Eq.~(\ref{modifiedFriedmann}). Setting $\dot{a}=0$
in~(\ref{modifiedFriedmann}) yields the  equation
locating the critical radius of the fluid sphere:
\begin{eqnarray}
1-\frac{8\pi}{3}\frac{\rho_f}{ka^f}
\left(1-\frac{\rho_f}{\rho_{cr}a^{f+2}}\right)=0\,,
\end{eqnarray}
or
\begin{eqnarray}
1-\frac{8\pi}{3}\left(\frac{r_1}{r_0}\right)^2
\frac{\rho_f \, r_0^{f+2}}{x^f}\left(1-\frac{
\rho_f \, r_0^{f+2}}{\rho_{cr}x^{f+2}}
\right)=0\,,\label{31}
\end{eqnarray}
where $k$ has the dimensions of the inverse of a squared
length, hence we define
\begin{eqnarray}
k\equiv\frac{1}{r_1^2}\,.
\end{eqnarray}
Eq.~(\ref{31}) provides two critical values of the radius
$x$ (a proper length or areal radius), the maximum
radius $x_\text{max} $ and the minimum radius
$x_\text{min}$ of
the
fluid sphere. If $r_1=r_0$ we have
\begin{eqnarray}
x_\text{min}=x_{-}\,, \ \ \ \  \  x_\text{max}=x_{+}\,,
\end{eqnarray}
which means that the minimum and maximum radii of the
sphere coincide with those of  the {inner and
outer apparent horizons}, respectively.

If instead $r_1 > r_0$ we have
\begin{eqnarray}
x_\text{min}<x_{-}\,, \ \ \ \  \  x_\text{max}>x_{+}\,:
\end{eqnarray}
the fluid sphere can pulsate across the
{outer apparent} horizon $x_+$
and the {inner apparent} horizon $x_{-}$.

If instead $r_1 < r_0$ we get
\begin{eqnarray}
x_\text{min} > x_{-}\,, \ \ \ \  \  x_\text{max} < x_{+}\,,
\end{eqnarray}
then the fluid sphere pulsates between the {two
apparent horizons $x_+$ and $x_{-}$.} In this case the
surface of infinite redshift $x_{+}$ conceals the sphere
from the observer at infinity, who still sees a black hole.

\begin{figure}[h]
\begin{center}
\includegraphics[width=9cm]{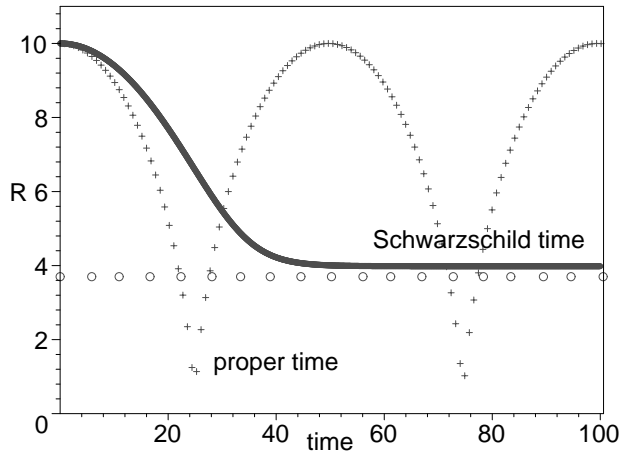}
\caption{The variation of the sphere surface with the proper time $t$ and
Schwarzschild time $\tilde{t}$,
starting at rest at $R_0=10$ and falling toward the center
(for the parameter values $f=1$, $M_1=2$, $l=0.79$, and
$E=0.77$).}\label{fig:2}
\end{center}
\end{figure}

We stress that the formation of an {apparent}
horizon is not avoided since we always have $x_\text{min} <
x_{+}$.
Secondly, the pulsation of the sphere between minimum
and maximum radii is experienced by
the comoving observer but not by the observer at
infinity, for whom the
sphere will take an infinite time to reach the
{outer apparent}
horizon. Therefore, the pulsation of the sphere
is completely unobservable by the observer at infinity. In
Fig.~\ref{fig:2} we plot the variation of proper time $t$
for the comoving observer and the Schwarzschild time
$\tilde{t}$
for the observer at infinity during the process of
collapse (or pulsation) of the sphere.{The variation of the sphere surface with the proper time $t$ and the Schwarzschild time $\tilde{t}$ are determined by Eq.~(\ref{45}) and}
\begin{eqnarray}
\frac{dR}{d\tilde{t}}&=&-\frac{E}{\sqrt{E^2-\left[1-\frac{2M_f}{R^{f}}\left(1-\frac{l^{f+2}}{R^{f+2}}\right)\right]}}\nonumber\\&&\cdot
\left[1-\frac{2M_f}{R^{f}}\left(1-\frac{l^{f+2}}{R^{f+2}}\right)\right]\,,
\end{eqnarray}
{respectively.}

For realistic astrophysical objects and the case  $f=1$,
Eq.~(\ref{31}) can be safely approximated by
\begin{eqnarray}
\left(\frac{r_0}{r_1}\right)^2=\frac{2M_1}{x}\,. \label{36}
\end{eqnarray}
The right hand side of Eq.~(\ref{36}) is  twice  the
negative of the Newtonian potential $-M_1/x$. In general,
we have
\begin{eqnarray}
\frac{2M_1}{x}\ll 1
\end{eqnarray}
and
\begin{eqnarray}
r_0\ll r_1\,.
\end{eqnarray}
In this case the fluid sphere can oscillate across the
{outer} horizon $x_{+}$. But for the observer at
infinity,
it takes an infinite time for the sphere to cross this
horizon. The case $r_1 < r_0$ is interesting: in this
case the fluid sphere oscillates inside the outer horizon
$x_+$, behaving as a quantum ``beating heart'' of the
black hole. To summarize, the black hole singularity
is now replaced by a pulsating sphere, the pulsation of
which is completely
unobservable by an observer at infinity.

We remind the reader that the density distribution within
the sphere
$r_0$ is assumed to be perfectly spherical. In
reality, this density distribution could be asymmetric and
the quadrupole moment be time-dependent. Then, presumably,
the energy loss due to gravitational wave radiation (if
still applicable) would damp
the oscillation and the beating of the ``quantum heart'',
and lead to $a_{\rm max} \rightarrow a_{\rm min}$.

In the next section we discuss the matching between
exterior and interior geometries at the
surface of the fluid sphere.

\section{Matching interior and exterior geometries}
\label{Sec:4}

The field equations are satisfied on the surface of the
sphere if and only if the induced 3-metrics and extrinsic
curvatures of the surface's three-dimensional world tube
are the same, whether measured from its interior or from
its exterior. In this section we verify that this is indeed
the case for the oscillating sphere discussed above.

The surface of the sphere has constant
radial coordinate $r=r_0$. Eq.~(\ref{1}) gives the
proper circumferential radius of the sphere
$R(t)=a(t) \, r_0$, where $a(t)$ obeys
Eq.~(\ref{modifiedFriedmann}).
Assuming that the fluid is initially at rest at time $t=0$,
it is $a(0)=a_0$, $\dot{a}(0)=0$ and
\begin{equation}
k=\frac{8\pi}{3}\frac{\rho_f }{a_0^f}\left(1-\frac{\rho_f
}{\rho_{cr} \, a_0^{f+2}}\right)\,.\label{40}
\end{equation}
Eq.~(\ref{modifiedFriedmann}) then assumes the form
\begin{equation}
\dot{a}^2=\frac{8\pi}{3}\left[\frac{\rho_f
}{a^f}\left(1-\frac{\rho_f
}{\rho_{cr}a^{f+2}}\right)-\frac{\rho_f
}{a_0^f}\left(1-\frac{\rho_f  }{\rho_{cr}
\, a_0^{f+2}}\right)\right]\, . \label{41}
\end{equation}
Once $a(t)$ is derived from Eq.~(\ref{41}), the intrinsic
3-metric induced on the surface $\Sigma$ of the sphere, as
measured
from
the FLRW interior, is obtained by setting $r=r_0$ in
Eq.~(\ref{1}):
\begin{equation}
ds^2_{(-)}=-dt^2+a^2 r_0^2d\Omega_{(2)}^2\,. \label{42}
\end{equation}
Let us focus now on the exterior metric~(\ref{22}). The
surface of the sphere is at coordinate
\begin{equation}
x=R\left({t}\right)=a(t) \, r_0\,, \label{43}
\end{equation}
and $t$ is the proper time of this surface. The 4-velocity
$v^a$ of this surface is tangent to its spacetime
trajectory, which is a timelike curve and, therefore,
satisfies $v^c v_c =-1$. This normalization determines
$R({t})$ through
\begin{equation}
\dot{R}^2=E^2-\left[1-\frac{2M_f}{R^{f}}
\left(1-\frac{l^{f+2}}{R^{f+2}}\right)\right]\,.\label{45}
\end{equation}
At $t=0$ and $R=R_0=a_0 r_0$, the radial velocity of the
surface vanishes, $dR/dt=0$, which gives
\begin{equation}
E^2= 1-\frac{2M_f}{R^{f}_0}
\left(1-\frac{l^{f+2}}{R^{f+2}_0}\right) \,.
\end{equation}
Eq.~(\ref{45}) then takes the form
\begin{equation}
\dot{R}^2=\frac{2M_f}{R^{f}}\left(1-\frac{l^{f+2}}{
R^{f+2}}\right)-\frac{2M_f}{R^{f}_0}\left(
1-\frac{l^{f+2}}{R^{f+2}_0}\right)\,. \label{47}
\end{equation}
Eqs.~(\ref{41}) and~(\ref{47}) agree if and
only if
\begin{equation}
R=ar_0\,,\ \ \ \ \ R_0=a_0r_0\,.\label{46}
\end{equation}
Substituting $x=R$ with Eq.~(\ref{45}) and Eq.~(\ref{46})
into the exterior solution~(\ref{22}), we find the line
element on the surface to be
\begin{eqnarray}
d s_{(+)}^2&=&-\left[1-\frac{2M_f}{R^{f}}
\left(1-\frac{l^{f+2}}{R^{f+2}}\right)\right]
\left(\frac{d\tilde{t}}{dt}\right)^2dt^2
+R^2d\Omega^2_{(2)}\nonumber\\
&&\nonumber\\
&&  +\left[1-\frac{2M_f}{R^{f}}\left(1-\frac{l^{f+2}}{
R^{f+2}}\right)\right]^{-1}
\left(\frac{dR}{dt}\right)^2dt^2\, \nonumber\\
&&\nonumber\\
&=&-\left[1-\frac{2M_f}{R^{f}}\left(
1-\frac{l^{f+2}}{R^{f+2}}\right)\right]^{-1}
E^2dt^2+R^2d\Omega_{(2)}^2\nonumber\\
&&\nonumber\\
&& +\left[1-\frac{2M_f}{R^{f}}\left(1-\frac{l^{f+2}}{
R^{f+2}}\right)\right]^{-1}\left(\frac{dR}{dt}\right)^2dt^2
\nonumber\\
&&\nonumber\\
& = &-dt^2+a^2 r_0^2d\Omega_{(2)}^2\,,
\end{eqnarray}
which is exactly the 3-metric~(\ref{41}) as measured from
the FLRW interior of the sphere. Therefore, the 3-metrics
induced by the exterior and interior 4-dimensional metrics
match {continuously} on the surface of the sphere.

Let us consider now the extrinsic curvatures of both
interior and exterior. It is necessary and
sufficient to show that the extrinsic curvatures $K_{ij}$
of the surface are the same whether measured in the
exterior or the interior. First, we calculate
$K_{ij}^{(-)}$ in the FLRW interior. Since $t$ is the
proper time on the surface $\Sigma$, the components of its
4-velocity in coordinates $\left( \tilde{t}, x, \theta,
\varphi \right)$ are
\begin{eqnarray}
v^{\mu}&=& \frac{dx^{\mu}}{dt}=\left(
\frac{d\tilde{t}}{dt},
\frac{dR}{dt}, 0, 0 \right) \nonumber\\
&&\nonumber\\
&=& \left(
\frac{E}{1-\frac{2M_f}{R^{f}}\left(1-\frac{l^{f+2}}{
R^{f+2}}\right) }, \dot{R}, 0,0 \right) \,.
\end{eqnarray}
The normal vector to $\Sigma$  is
\begin{equation}
n^a=\frac{\sqrt{1-kr_0^2}}{a}
\left( \frac{\partial}{\partial
r} \right)^a \,,
\end{equation}
while the vectors $v^a =\left( \partial/\partial
t\right)^a ,\ \left( \partial/\partial\theta\right)^a $, and $
\left( \partial/\partial\phi \right)^a$ lie in the
surface. Let the indices $i$ and $j$ run over $t,\
\theta$, and $\phi$.  Then we have
\begin{equation}
 K_{ij}^{(-)} \equiv e_{(i)}^a
e_{(j)}^b  \nabla_{a} n_b =\frac{1}{2} \,\pounds_{n}
g_{ij}\,,\label{50}
\end{equation}
where $\left\{ e^a_{(t)},  e^a_{(\theta)},  e^a_{(\varphi)}
\right\}=\left\{  e^a_{(i)} \right\}$ is a triad of vectors
on the surface of the sphere, the subscript $(i)$ denotes a
triad index, and $\pounds_n$ is the Lie
derivative along the direction of the unit normal $n^a$.
The metric~(\ref{1}) gives
\begin{eqnarray}
&& K_{tt}^{(-)}=K_{t\theta}^{(-)}=
K_{t\phi}^{(-)}=K_{\theta\phi}^{(-)}=0\,, \\
&&\nonumber\\
&& K_{\theta\theta}^{(-)}=K_{\phi\phi}^{(-)}
\sin^{-2}\theta=ar_0\sqrt{1-k r_0^2}\, .
\end{eqnarray}
We then calculate the extrinsic curvature
$K_{ij}^{(+)}$ in the exterior region. In the background of
the exterior spacetime~(\ref{22}), the 4-velocity of the
spherical surface is
\begin{eqnarray}
v^a = v^{\tilde{t}}e_{(\tilde{t})}^a + v^{x} e_{(x)}^a\,.
\end{eqnarray}
The normal vector is
\begin{eqnarray}
n^a =n^{\tilde{t}} e_{(\tilde{t})}^a + n^{x} e_{(x)}^a \,.
\end{eqnarray}
These vectors satisfy the conditions
\begin{eqnarray}
n^c  n_c =1\,,\ \ \ n^c  v_c =0\,,\ \ \ v^c  v_c =-1\,,
\end{eqnarray}
from which one obtains also
\begin{eqnarray}
n^{\tilde{t}}=v_{x}\,,\ \ \ n^{x}=-v_{\tilde{t}}\,.
\end{eqnarray}
Let $i$ and $j$ run over the values $t,\ \theta$, and
$\phi$ as before, then Eq.~(\ref{50}) holds for the
exterior
metric, with
\begin{eqnarray}
e_{(\tilde{t})}^c  e^{(\tilde{t})}_c &=& v^c  v_c=-1\,,\\
&&\nonumber\\
e_{(\tilde{t})}^c e^{(\theta)}_c &=&  e_{(\tilde{t})}^c
e^{(\phi)}_c =e_{(\theta)}^c  e^{(\phi)}_c=0\,.
\end{eqnarray}
We have
\begin{eqnarray}
K_{{t}{t}}^{(+)}=
K_{{t}\theta}^{(+)}=K_{{t}\phi}^{(+)}=
K_{\theta\phi}^{(+)}=0\;,
\end{eqnarray}
and
\begin{eqnarray}
&&K_{\theta\theta}^{(+)}=
K_{\phi\phi}^{(+)}\sin^{-2}\theta=
\frac{1}{2}\left(x^2\right)_{,n}\nonumber\\
&&\nonumber\\
&&=\frac{1}{2}\left(x^2\right)_{,x}n^{x}
 =-x v_{\tilde{t}} =RE \nonumber\\
&&\nonumber\\
&& =R\left[1-\frac{2M_f}{R^{f}_0}\left(
1-\frac{l^{f+2}}{R^{f+2}_0}\right)\right]^{1/2}
\nonumber\\
&&\nonumber\\
&&=ar_0\sqrt{1-k r_0^2}\,,
\end{eqnarray}
where we used Eqs.~(\ref{42}),~(\ref{46}), and
\begin{eqnarray}
 v^{\tilde{t}}=\frac{d\tilde{t}}{dt}=
\frac{\left[1-\frac{2M_f}{R^{f}_0}
 \left(1-\frac{l^{f+2}}{R^{f+2}_0}\right)\right]^{
\frac{1}{2}}}{1-\frac{2M_f}{R^{f}}\left(
1-\frac{l^{f+2}}{R^{f+2}}\right)} \,.
\end{eqnarray}
Therefore, it is
\begin{eqnarray}
 K_{ij}^{(-)}=K_{ij}^{(+)}\,,
\end{eqnarray}
which completes the proof.

\section{Discussion and conclusions}
\label{Sec:5}

The Hawking-Penrose singularity theorem, based on general relativity,
dictates that there is a singularity at the center of a black hole.
In order to remove this
singularity, one must resort to a more fundamental quantum theory of
gravity. We currently have two main contenders to the role of such a
theory, LQG and string theory. Using the dynamical equation arising in
LQC or braneworld models, we have studied the gravitational collapse of
a perfect fluid sphere. \emph{Within the context of LQC, note here that an approximation
to the full effective dynamics is adopted in the present work. This
approximation underlies the replacement of the full
Eq.~(\ref{modifiedFriedmann1}) with Eq.~(\ref{modifiedFriedmann}) at the
beginning of our discussion. }

We find that, in the comoving frame, the sphere does not
collapse to a singularity but pulsates instead between a
maximum and a minimum size, effectively removing the
singularity from
the gravitational collapse. In the process of seeking
the solution exterior to the sphere, we propose a method
for constructing the exterior solution.
We then find that the exterior solution usually has two
horizons, which is reminiscent of the Reissner-Nordstr\"om
black hole spacetime of classical relativity. As a result,
in the frame of an observer at spatial infinity, the
collapsing fluid takes an infinite coordinate time to cross
the {outer} horizon and the pulsations of the
quantum-corrected core are completely unobservable by the
far-away observer. Borrowing current terminology from black
hole astrophysics, the pulsating core hidden inside the
{apparent} horizon plays the role of a ``beating
heart'' for the
black hole.

\section*{Acknowledgments}

{We thank the referee for useful suggestions and for
bringing several references to our attention.} This work is
partially supported by the Strategic Priority Research
Program ``Multi-wavelength Gravitational Wave Universe'' of the
CAS, Grant No. XDB23040100, and the NSFC
under grants 10973014, 11373020, 11465012, 11633004 and 11690024, and
the Project of CAS, QYZDJ-SSW-SLH017. VF thanks the Natural
Science and Engineering Research Council of Canada for
partial support.

\appendix

\section{From interior to exterior solution}
\label{AppendixA}

Here we show how to obtain the exterior
solution from the interior one using our ``surface
trick'' in the context of general relativity.

The metric for a homogenous and isotropic fluid sphere is
given by Eq.~(\ref{1}), where the scale factor $a$ obeys
Eq.~(\ref{Friedmann}). Following the
procedure of Sec.~\ref{Sec:3}, we rewrite Eq.~(\ref{1})
in the frame of the observer at infinity as
\begin{eqnarray}
ds^2&=&-F^2\left(1-\frac{8\pi}{3}\rho
x^2\right)d\bar{t}^2+x^2d\Omega_{(2)}^2\nonumber\\
&&\nonumber\\
&& +\left(1-\frac{8\pi}{3}\rho x^2\right)^{-1}dx^2\,,
\end{eqnarray}
where the metric components must be understood as
functions of $\bar{t}$ and $x$.

At the surface $r=r_0$ of the sphere ($r_0$ is a constant,
since $r$ is a comoving coordinate \cite{MTW:1973}), we
have
\begin{eqnarray}
ds^2&=&-\left[1-\frac{8\pi}{3}\rho
\left(\frac{x}{r_0}\right) x^2\right]d\tilde{t}^2
+x^2d\Omega_{(2)}^2\nonumber\\
&&\nonumber\\
&& +\left[1-\frac{8\pi}{3}\rho\left(\frac{x}{r_0}\right)
x^2\right]^{-1}dx^2\,.\label{A2}
\end{eqnarray}
The metric~(\ref{A2}) also represents the exterior
metric at
the surface $r=r_0$. Thus we can take this static
spherically symmetric spacetime as the exterior solution of
the fluid sphere. Similar to the proof in
Sec.~\ref{Sec:4}, we
can show that the exterior geometry~(\ref{A2}) matches
{continuously} the interior one~(\ref{1}).

As an example, we consider the energy density
\begin{eqnarray}
\rho=\lambda+\frac{\rho_m}{a^3}+\frac{\rho_r}{a^4}\,,
\label{A3}
\end{eqnarray}
with $\lambda$ for the cosmological constant energy
density, ${\rho_m}/{a^3}$ for the dust density, and
${\rho_r}/{a^4}$ for the radiation density. Substituting
Eq.~(\ref{A3}) with
$a=x/r_0$ into Eq.~(\ref{A2}), one obtains
\begin{eqnarray}
ds^2&=&-\left(1-\frac{2M}{x}-\frac{Q^2}{x^2} -\Lambda
x^2\right)d\tilde{t}^2+x^2d\Omega_{(2)}^2\nonumber\\
&&\nonumber\\
&& +\left(1-\frac{2M}{x}-\frac{Q^2}{x^2}-\Lambda
x^2\right)^{-1}dx^2\,,\label{A4}
\end{eqnarray}
where
\begin{eqnarray}
M\equiv \frac{4\pi}{3}\rho_m r_0^{3}\,,\ \ \ \ \
Q^2\equiv\frac{8\pi}{3}\rho_r  r_0^{4}\,,\ \ \ \ \
\Lambda\equiv\frac{8\pi}{3}\lambda\,.
\end{eqnarray}
The metric~(\ref{A4}) is the
Reissner-Nordstr\"om-de Sitter solution of the Einstein
equations with the formal replacement $Q\rightarrow i Q$.
While $M$ plays the role of a mass, $Q^2$ also plays the
role of the mass of radiation (not of electric charge)
contained in a sphere. In fact, the thermal bath of
radiation with density $\rho_r/a^4$ has nothing to do with
free electric charges and the analogy with
Reissner-Nordstr\"om-de Sitter is purely formal and not
complete (because of the opposite sign of the term
in $Q^2/x^2$).

\section{From exterior to interior solution}
\label{AppendixB}

Here we show how to obtain the interior geometry from the
exterior one in the context of general
relativity. We focus on the static spherically symmetric
spacetime
\begin{eqnarray}
ds^2&=&-\left[1-\frac{8\pi}{3}
\rho\left(x\right)x^2\right]d\tilde{t}^2+x^2d\Omega_{(2)}^2
\nonumber\\
&&\nonumber\\
&& +\left[1-\frac{8\pi}{3}\rho\left(
x\right)x^2\right]^{-1}dx^2\,.\label{B1}
\end{eqnarray}
We assume that the spacetime exterior to a perfect
fluid sphere is given by Eq.~(\ref{B1}). Then our task is
to look for
the interior solution starting from this exterior.
The normalization of the tangent $v^a$ to the
trajectory described by the surface of the sphere in
spacetime gives
\begin{eqnarray}
\dot{x}^2=E^2-1+\frac{8\pi}{3}\rho x^2\,,\label{B2}
\end{eqnarray}
with an overdot denoting differentiation with respect to
the proper time $t$ and where $E$ is an integration
constant. Eqs.~(\ref{B2}) and~(\ref{Friedmann}) coincide
provided that
\begin{eqnarray}
x=a r_0\,\ \ \ \ \ \ -k r_0^2=E^2-1\,,\label{B3}
\end{eqnarray}
with $r_0$ a constant. This coincidence motivates us to
consider simply the homogenous and isotropic perfect fluid
sphere as the interior solution. Then $\rho$ plays
the role of the energy density in the comoving frame.  We
have checked that the resulting interior
solution does match the exterior solution
{continuously}. In the
following we consider, as an example, the Bardeen
metric \cite{bard:1968} as the exterior geometry outside a
perfect fluid sphere.

The Bardeen line element describing a regular black hole
is
\begin{eqnarray}
ds^2&=&-\left[1-\frac{2Mx^2}{
\left(x^2+q^2\right)^{3/2}}\right]d\tilde{t}^2
+x^2d\Omega_{(2)}^2\nonumber\\
&&\nonumber\\
&& +\left[1-\frac{2Mx^2}{\left(x^2+q^2\right)^{3/2}}
\right]^{-1}dx^2\,,\label{B4}
\end{eqnarray}
where $M$ is the mass and $q$ is the monopole charge of a
self-gravitating magnetic field described by a nonlinear
electrodynamics \cite{beato:2000}. This model has been
revisited by Borde, who clarified the avoidance of
singularities in this spacetime
\cite{borde:1994,borde:1997}. For a certain range of the
parameter $q$, the Bardeen metric describes a black
hole. When $x\gg q$, it behaves as the Schwarzschild black
hole but, when $x\ll q$, it behaves as de Sitter
space, therefore, the spacetime in general has two
{apparent} horizons. Explicitly, there are two
{such} horizons when
\begin{eqnarray}
|q|<\frac{4M}{3\sqrt{3}} \,.
\end{eqnarray}
Given the exterior metric~(\ref{B4}), one can read
off the corresponding energy density of the perfect fluid
sphere, which can be parameterized as
\begin{eqnarray}
\rho&=&\frac{\rho_m
}{\left(a^2+a_0^2\right)^{3/2}}\,,\label{B6}
\end{eqnarray}
where $\rho_m$ and $a_0$ are two constants and $a(t)$ is
the time-dependent  radius of the fluid sphere.
Substituting Eq.~(\ref{B6}) into Eq.~(\ref{Friedmann})
yields
\begin{equation}
{\dot{a}^2}=\frac{8\pi}{3}\frac{\rho_m a^2}{\left(
a^2+a_0^2\right)^{3/2}}-k\,,
\end{equation}
describing a pulsating sphere. However, the pulsation is
again unobservable by an observer at infinity due to the
unavoidable formation of an {apparent} horizon. On the other hand, if we regard
the above equation as the Friedmann equation to study its cosmic evolution, we
find it can give an exponential inflationary universe provided that $a\ll a_0$.
Thus it would be interesting to investigate this inflationary model in great detail.

\newcommand\ARNPS[3]{~Ann. Rev. Nucl. Part. Sci.{\bf ~#1}, #2~ (#3)}
\newcommand\AL[3]{~Astron. Lett.{\bf ~#1}, #2~ (#3)}
\newcommand\AP[3]{~Astropart. Phys.{\bf ~#1}, #2~ (#3)}
\newcommand\AJ[3]{~Astron. J.{\bf ~#1}, #2~(#3)}
\newcommand\APJ[3]{~Astrophys. J.{\bf ~#1}, #2~ (#3)}
\newcommand\APJL[3]{~Astrophys. J. Lett. {\bf ~#1}, L#2~(#3)}
\newcommand\APJS[3]{~Astrophys. J. Suppl. Ser.{\bf ~#1}, #2~(#3)}
\newcommand\JHEP[3]{~JHEP.{\bf ~#1}, #2~(#3)}
\newcommand\JMP[3]{~J. Math. Phys. {\bf ~#1}, #2~(#3)}
\newcommand\JCAP[3]{~JCAP {\bf ~#1}, #2~ (#3)}
\newcommand\LRR[3]{~Living Rev. Relativity. {\bf ~#1}, #2~ (#3)}
\newcommand\MNRAS[3]{~Mon. Not. R. Astron. Soc.{\bf ~#1}, #2~(#3)}
\newcommand\MNRASL[3]{~Mon. Not. R. Astron. Soc.{\bf ~#1}, L#2~(#3)}
\newcommand\NPB[3]{~Nucl. Phys. B{\bf ~#1}, #2~(#3)}
\newcommand\CMP[3]{~Comm. Math. Phys.{\bf ~#1}, #2~(#3)}
\newcommand\CQG[3]{~Class. Quantum Grav.{\bf ~#1}, #2~(#3)}
\newcommand\PLB[3]{~Phys. Lett. B{\bf ~#1}, #2~(#3)}
\newcommand\PRL[3]{~Phys. Rev. Lett.{\bf ~#1}, #2~(#3)}
\newcommand\PR[3]{~Phys. Rep.{\bf ~#1}, #2~(#3)}
\newcommand\PRd[3]{~Phys. Rev.{\bf ~#1}, #2~(#3)}
\newcommand\PRD[3]{~Phys. Rev. D{\bf ~#1}, #2~(#3)}
\newcommand\RMP[3]{~Rev. Mod. Phys.{\bf ~#1}, #2~(#3)}
\newcommand\SJNP[3]{~Sov. J. Nucl. Phys.{\bf ~#1}, #2~(#3)}
\newcommand\ZPC[3]{~Z. Phys. C{\bf ~#1}, #2~(#3)}
\newcommand\IJGMP[3]{~Int. J. Geom. Meth. Mod. Phys.{\bf ~#1}, #2~(#3)}
\newcommand\GRG[3]{~Gen. Rel. Grav.{\bf ~#1}, #2~(#3)}
\newcommand\EPJC[3]{~Eur. Phys. J. C{\bf ~#1}, #2~(#3)}
\newcommand\PRSLA[3]{~Proc. Roy. Soc. Lond. A {\bf ~#1}, #2~(#3)}
\newcommand\AHEP[3]{~Adv. High Energy Phys.{\bf ~#1}, #2~(#3)}
\newcommand\Pramana[3]{~Pramana.{\bf ~#1}, #2~(#3)}


\begin{thebibliography}{99}

\bibitem{haw:1970} S.W. Hawking and R. Penrose,
\PRSLA{314}{529}{1970}.

\bibitem{sch:2003}J.H. Schwarz, in {\em Measuring and
Modeling the Universe}, Proceedings, Pasadena, USA,
November 17-22, 2002 W.L. Freedman ed. (Cambridge
University Press, Cambridge, UK 2004), pp.~53-66
[arXiv:astro-ph/0304507].

\bibitem{lqg:2004}See, {\em e.g.}, A. Ashtekar and J.
Lewandowski, \CQG{21}{R53}{2004}.

\bibitem{open:1939}J.R. Oppenheimer and H. Snyder,
\PRD{56}{455}{1939}.

\bibitem{AFW03} A. Ashtekar, S. Fairhurst, and J.L. Willis,
Class. Quantum Grav. {\bf 20}, 1031 (2003).

\bibitem{Boe:2007} C.G. Boehmer and K. Vandersloot,
\PRD{76}{104030}{2007} [arXiv:0709.2129].

\bibitem{mod:2006} L. Modesto, \CQG{23}{5587}{2006}
[arXiv:gr-qc/0509078].

\bibitem{ale:2015}A. Corichi and P. Singh, \CQG{33}{055006}{2016} [arXiv:gr-qc/1506.08015].

\bibitem{ol:2017}J. Olmedo, S. Saini, and P. Singh,
arXiv:gr-qc/1707.07333.

\bibitem{cam:2007} M. Campiglia, R. Gambini, and J. Pullin,
\CQG{24}{3649}{2007} [arXiv:gr-qc/0703135].

\bibitem{boj:2005} M. Bojowald, R. Goswami, R. Maartens,
and P. Singh, \PRL{95}{091302}{2005}.

\bibitem{mod:2008} L. Modesto, \AHEP{2008}{459290}{2008}.

\bibitem{chiou:2008} D.W. Chiou, \PRD{78}{064040}{2008}
[arXiv:gr-qc/0611043].

\bibitem{boj:2008} M. Bojowald, T. Harada, and R.
Tibrewala,\PRD{78}{064057}{2008}.

\bibitem{Rod:2008} G. Rodolfo and J. Pullin, \PRL{101}{161301}{2008}.

\bibitem{TippettHusain11} B.K. Tippett and V. Husain,
Phys. Rev. D {\bf 84}, 104031 (2011).

\bibitem{ModestoPRD04} L. Modesto, Phys. Rev. D {\bf
70}, 124009 (2004).

\bibitem{ZiprickKunstatter} J. Ziprick and G. Kunstatter,
Phys. Rev. D {\bf 80}, 024032 (2009).

\bibitem{PeltolaKunstatter} A. Peltola and G. Kunstatter,
Phys. Rev. D {\bf 80}, 044031 (2009).


\bibitem{bambi}C. Bambi, D. Malafarina, and L. Modesto,
Phys. Rev. D {\bf 88}, 044009(2013).

\bibitem{liu}Y. Liu, D. Malafarina, L. Modesto, and C.
Bambi, Phys. Rev. D {\bf 90}, 044040 (2014).

\bibitem{mala}D. Malafarina, Universe {\bf 3}, 48 (2017).

\bibitem{Waldbook} R.M. Wald, {\em General Relativity}
(Chicago University Press, Chicago, 1984).

\bibitem{haw:1973} S.W. Hawking and G.F.R. Ellis, \emph{The
Large Scale Structure of Space-Time} (Cambridge University
Press, Cambridge, UK 1973), p.~135.

\bibitem{lqc:001}G. Vereshchagin, \JCAP{0407}{013}{2004}
[arXiv:gr-qc/0406108].

\bibitem{lqc:002} P. Singh, K. Vandersloot, and G.
Vereshchagin, \PRD{74}{043510}{2006} [arXiv:gr-qc/0606032].

\bibitem{lqc:003}A. Ashtekar, T. Pawlowski, P. Singh, and
K. Vandersloot, \PRD{75}{024035}{2007}
[arXiv:gr-qc/0612104].


\bibitem{lqc:0031}P. Singh and F. Vidotto, \PRD{83}{064027}{2011} [arXiv:gr-qc/1012.1307].


\bibitem{lqc:0032}J. L. Dupuy and P. Singh, \PRD{95}{023510}{2017}
[arXiv:gr-qc/1608.07772].






\bibitem{lqc:004} A. Ashtekar, T. Pawlowski, and P. Singh,
\PRD{74}{084003}{2006} [arXiv:gr-qc/0607039].

\bibitem{para:14}P. Diener, B. Gupt and P. Singh, \CQG{31}{105015}{2014} [arXiv:gr-qc/1402.6613].



\bibitem{lqc:005}A. Ashtekar, T. Pawlowski, and P. Singh,
\PRL{96}{141301}{2006} [arXiv:gr-qc/0602086].

\bibitem{lqc:006}E. Wilson-Ewing, \JCAP{1303}{026}{2013}
[arXiv:1211.6269].




\bibitem{Bojowaldbook} M. Bojowald, {\em Quantum Cosmology}
(Springer, New York, 2011).

\bibitem{bw:001}Y. Shtanov and V. Sahni, \PLB{557}{1}{2003}
[arXiv:gr-qc/0208047].

\bibitem{bw:002}M. Seikel and M. Camenzind, \PRD{79}{083531}{2009} [arXiv:astro-ph/0811.4629].

\bibitem{bw:003} M.G. Brown, K. Freese, and W.H. Kinney,
\JCAP{0803}{002}{2008} [arXiv:astro-ph/0405353].



\bibitem{lqc:007} A. Ashtekar, AIP Conf. Proc. {\bf 861}, 3
(2006) [arXiv:gr-qc/0605011].


\bibitem{lqc:0071} G. J. Olmo and  P. Singh,
\JCAP{0901}{030}{2009} [arXiv:gr-qc/0806.2783].

\bibitem{rev:2} A. Corichi and E. Montoya,
\PRD{84}{044021}{2011}.

\bibitem{piter:14}P. Diener, B. Gupt, M. Megevand, and P.
Singh, \CQG{31}{165006}{2014} [arXiv:hep-ph/1406.1486].


\bibitem{rev:3} C. Rovelli {and E.} Wilson-Ewing,
\PRD{90}{023538}{2014}.

\bibitem{wein:1972} S. Weinberg, {\em Gravitation and
Cosmology} (Wiley, New York, 1972).


\bibitem{wang:2013} R. Y. Yu and T. Wang, \Pramana{80}{349}{2013}; T. Wang, \CQG{32}{195006}{2015}.


\bibitem{MTW:1973} C.W. Misner, K.S. Thorne, and J.A.
Wheeler, {\em Gravitation} (San Francisco, Freeman,
1973).

\bibitem{Israel} W. Israel, Nuovo Cimento~B {\bf 44}, 1
(1966); {\em Erratum} {\bf 48}, 463 (1967).

\bibitem{Poisson} E. Poisson, {\em A Relativist's Toolkit}
(Cambridge University Press, Cambridge, UK 2004).

\bibitem{barcelo:2016} C. Barcel\'{o}, R. Carballo-Rubio,
and L. J. Garay, Universe {\bf2}, 7 (2016).

\bibitem{kis:2003} V.V. Kiselev, \CQG{20}{1187}{2003}.

\bibitem{abe:2016} J. Abedi and H. Arfaei,
\JHEP{03}{135}{2016}.

\bibitem{chen:2015}P. Chen, Y. C. Ong, and D.-h. Yeom,
\PR{603}{1}{2015}.

\bibitem{bard:1968} J.M. Bardeen, in Conference Proceedings
of GR5 (Tbilisi, USSR, 1968), p.~174.

\bibitem{beato:2000} E. Ay\'on-Beato and A. Garcia,
\PLB{493}{149}{2000}.

\bibitem{borde:1994} A. Borde, \PRD{50}{3392}{1994}.

\bibitem{borde:1997} A. Borde, \PRD{55}{7615}{1997}.

\end{thebibliography}
\end{document}